\documentclass[12pt,preprint]{aastex}

\slugcomment{To appear in $The$ $Astrophysical$ $Journal$}

\shorttitle{The abundance robustness of the r-process stars}
\shortauthors{Niu et al.}

\begin{document}
\title {STUDY OF THE ELEMENT ABUNDANCES IN HD 140283:
THE ABUNDANCE ROBUSTNESS OF THE WEAK r- AND MAIN r-PROCESS STARS}
\author{Ping Niu\altaffilmark{1,2}, Wenyuan Cui\altaffilmark{1}, Bo Zhang\altaffilmark{1,3}}

\affil{1.Department of Physics, Hebei Normal University, 20 Nanerhuan
Dong Road, Shijiazhuang 050024, China \\
2.Department of Physics, Shijiazhuang  University, Shijiazhuang
050035, China}

\altaffiltext{3}{Corresponding author. E-mail address:
zhangbo@mail.hebtu.edu.cn}

\begin{abstract}

Many works engaged to investigate the astrophysical origin of the
neutron-capture elements in the metal-poor star HD 140283. However,
no definite conclusions have been drawn. In this work, using the
abundance-decomposed approach, we find that the metal-poor star HD
140283 is a weak r-process star. Although this star is a weak
r-process star, its Ba abundance mainly originate from the main
r-process. This is the reason that the ratio [Ba/Eu]$=-0.58\pm 0.15$
for HD 140283 is close to the ratio of the main r-process. Based on
the comparison of the abundances in the six weak r-process stars, we
find that their element abundances possess robust nature. On the
other hand, we find that abundance robust nature of the extreme main
r-process stars ([r/Fe]$\geq$1.5) can be extended to the lighter
neutron-capture elements. Furthermore, the abundance characteristics
of the weak r-process and main r-process are investigated. The
abundance robustness of the two category r-process stars could be
used as the constraint of the r-process theory and be used to
investigate the astrophysical origins of the elements in the
metal-poor stars and population I stars.

\end{abstract}


\section{INTRODUCTION}

HD 140283 is a metal-poor star with metallicity [Fe/H]=-2.63 and
locate in the solar neighborhood. Historically, HD 140283 was paid
attention because of its low Ca and Fe abundances relative to
corresponding abundances in the Solar system \citep{Cha51}. The
neutron-capture elements, which are heavier than Fe-peak elements,
are produced by the r-process and the s-process \citep{Bur57}.
According to the s-process calculations, \cite{Arl99} reported that
the abundance fraction of the odd isotopes of Ba
($f_{odd}=[N(^{135}Ba)+N(^{137}Ba)]/N(Ba)$) produced by the
s-process is obviously different from the fraction produced by
r-process. The fractions of the s-process and r-process in the Solar
system are $f^{s}_{odd}=0.11$ and $f^{r}_{odd}=0.46$, respectively.
In this case, the fraction should be a important indicator to reveal
the contributions of the s- and r- process to the abundances of
neutron-capture elements of the metal-poor star. On the other hand,
the abundance ratios [Ba/Eu] of pure s-process and pure r-process in
the Solar system are 1.13 and -0.69, respectively \citep{Arl99}. The
observed abundance ratio of the metal-poor star is also an important
constraint to investigate the relative contributions of the
s-process and r-process.

Based on the spectrum analysis of HD 140283, \cite{Mag95} derived
the fraction $f_{odd}=0.06\pm 0.06$ and suggested that Ba abundance
come from the s-process. However, \cite{Lam02} obtained spectrum of
HD 140283 and derived $f_{odd}=0.30\pm 0.21$, which is close to the
fraction of the r-process. Furthermore, \cite{Gal10} found that the
fractions of more metal-poor stars, including HD 140283, are lower
than 0.2 and suggested that their results are not agreement with
current viewpoint that the neutron-capture elements of the
metal-poor star dominantly come from the r-process. Recently,
\cite{Siq12} derived the Eu abundance of HD 140283 and obtained the
abundance ratio [Ba/Eu]$=-0.58\pm 0.15$. Their results imply that Ba
and Eu elements in HD 140283 mainly originate from the r-process.
Obviously, the astrophysical origin of the elements in HD 140283 is
a debate issue up to now. Because of low abundance ratios of the
heavy elements relative to Fe-peak elements and very old age, it is
very important to investigate the astrophysical origins of the
elements in HD 140283 \cite{Bon13}.

For exploring the astrophysical origins of the elements in the
metal-poor star HD 140283, in Section 2, we calculate the relative
contributions from the weak r-process and main r-process to the
abundances of HD 140283. Furthermore, the robust nature of
abundances in the weak r-process stars and the main r-process stars
are investigated in Section 3. The discussion about abundance
characteristics of the main r-process and weak r-process are written
in Section 4. The conclusions are presented in Section 5.

\section{ASTROPHYSICAL ORIGIN OF MENTAL-POOR STAR HD 140283}

For exploring the element origins in the metal-poor star HD 140283,
we compare the observed abundances \citep{Hon04,Roe12,Siq12,Bon13}
with the abundances of the weak r- and main r-process. The abundance
of the ith element can be expressed as \citep{Li13}:

\begin{equation}
N_{i}(Z)=(C_{r,m}N_{i,r,m}+C_{r,w}N_{i,r,w})\times{10^{[Fe/H]}}
\end{equation}

where $N_{i,r,m}$ and $N_{i,r,w}$ are the abundances of the main
r-process and weak r-process respectively, which are adopted from Li
et al. (2013). $C_{r,m}$ and $C_{r,w}$ are the component
coefficients, which represent the relative contributions of the main
r-process and weak r-process to the elemental abundances of the
star. We adopt equation (1) to fit the observed abundances of HD
140283 by looking for the minimum reduced $\chi^{2}$. The derived
component coefficients are $C_{r,m}=0.6$ and $C_{r,w}=4.4$. The
calculated results are plotted in the Figure 1 by solid line and the
observed abundances of HD 140283 are shown by filled circles. From
the figure we can see that the calculations are consistent with the
observations.

As suggested by \cite{Li13}, one can find the star with particular
abundance pattern of the neutron-capture elements based on the
component coefficients. If the component coefficient of the weak
r-process is much larger than that of the main r-process, this star
should be formed in a gas cloud in which the neutron-capture
elements are dominantly originate from the weak r-process. Taking
the weak r-process stars the HD 122563 and HD 88609 as examples,
their component coefficients of the weak r-process are larger than
3.8 and the component coefficients of the main r-process are smaller
than 0.6 \citep{Li13}. On the other hand, If the component
coefficient of the main r-process is much larger than that of the
weak r-process, this star should be formed in a gas cloud in which
the neutron-capture elements are dominantly originate from the main
r-process. The component coefficients of the main r-process for the
main r-process stars CS 22892-052 and CS 31082-001 are larger than
50.0 and the component coefficients of the weak r-process for the
two stars are about 4.0 \citep{Li13}. The component coefficients of
HD 140283 are $C_{r,m}=0.6$ and $C_{r,w}=4.4$, which means that this
star is a weak r-process star. In order to investigate the
astrophysical origins of the abundances in HD 140283, Figure 2 shows
the comparisons of the calculated abundances of the main r-component
and weak r-component with the observed abundances. Although this
star is a weak r-process star, from the figure we can see that
element Ba mainly originate from the main r-process. This is the
reason that the ratio [Ba/Eu]$=-0.58\pm 0.15$ for HD 140283 is close
to the ratio of the main r-process.

The adopted approach is based on the observed abundances of HD
140283, so the related uncertainties should be contained in the
model parameters. In order to investigate the uncertainties of the
component coefficients, taking $C_{r,w} =4.4$, the top panel of
Figure 3 shows the calculated abundances $\log\varepsilon(Eu)$ as a
function of the component coefficient $C_{r,m}$. In this figure,
solid curves represent the calculated results. Dashed line
represents the observed abundance with error expressed by dotted
lines. The shaded region shows the allowed range of the component
coefficient. From the figure we can see that, there is a range of
the component coefficients, $C_{r,m} =0.6^{+0.3}_{-0.2}$ ,  in which
the calculated abundances fall into observed limits of
$\log\varepsilon(Eu)$ . The bottom panel expresses the reduced
$\chi^{2}$ as the function of $C_{r,m}$ and shows a minimum
$\chi^{2}$= 1.45 at $C_{r,m}$ = 0.6. The calculated results
illustrate that the component coefficient is constrained well.
Adopting similar approach, we derive the range of another component
coefficient: $C_{r,w} =4.4^{+0.3}_{-1.2}$.

\section{ ABUNDANCE ROBUSTNESS OF THE WEAK R-STARS AND MAIN R-STARS}

Because HD 122563 and HD 88609 are the weak r-process stars, their
abundances are taken as the representative for exploring the
characters of the weak r-process \citep{Mon07,Han14}. However, it is
unclear whether the abundance patterns of the weak r-process stars
are uniform. After adding HD 140283, six weak r-process stars have
been reported up to now. For exploring the abundance robustness of
the six weak r-process stars, firstly we scaled the observed
abundances of HD 88609 \citep{Hon04,Hon07}, BD +4$^{0}$2621
\citep{Joh02}, HD 4306 \citep{Hon04}, HD 237846 \citep{Roe10} and HD
140283 \citep{Hon04,Roe12,Siq12,Bon13} to those of HD 122563
\citep{Hon04,Hon07} and show their abundances in Figure 4 (a). The
figure shows that the abundance patterns of the five stars match the
abundance pattern of HD 122563 quite well within the observed
uncertainties. Furthermore, we derive the average abundance of the
six stars and scale the element abundances of the six stars to the
average abundances. In the Figure 4 (b), the observed abundances of
six weak r-process stars and the average abundances, which have been
vertically moved for the sake of display, are presented by filled
circles and solid lines. In the Figure 4 (c), the abundance offsets
($\bigtriangleup\log\varepsilon=\log\varepsilon(X)_{ave}-\log\varepsilon(X)_{obs}$)
of the six sample stars are shown. The Figure 4 (d) shows the rms
abundance offsets. From the figure we find that the observed
abundances of light elements, Fe-peak elements and neutron-capture
elements of the six stars match the averaged abundances within 0.2
dex for the most elements. The results also mean that the abundances
of light elements, Fe-peak elements and neutron-capture elements of
the six stars are in agreement with each other and indicate that the
abundance patterns of the weak r-process stars are uniform, which
should be important for investigating the site (or sites) and
associated physical conditions of the weak r-process
nucleosynthesis.

Because of the extreme overabundance of heavier neutron-capture
elements ($Z\geq56$) ([Eu/Fe]$>1.6$), two main r-process stars CS
22892-052 and CS 31082-001 were paid attention frequently. It is
interesting that the abundances of the heavier neutron-capture
elements of the two stars are agreement with the r-process
abundances of the Solar system. However, the abundances of the
lighter neutron-capture elements (37$\leq Z \leq$47) are lower than
the r-process abundances \citep{Cow99,Hil02,Sne08}. \cite{Zha10}
have found that the abundances of the r-process elements, including
lighter neutron-capture elements, of the Solar system could be
explained by the contributions of main r-process and weak r-process.
For understanding of the r-process nucleosynthesis, the abundance
uniform of the lighter neutron-capture elements in the main
r-process stars should be important. This reason inspire us to
explore the robustness of the abundances of the main r-process
stars. In this case, we specially pay attention to the abundances of
the neutron-capture elements in the extreme main r-process stars
([r/Fe]$\geq1.5$). Their heavy-element abundances are expected to be
dominantly originate from the main r-process. Six stars, CS
22892-052 \citep{Sne03}, CS 31082-001 \citep{Hil02}, CS 29497-004
\citep{Bar05}, HE 1219-0312 \citep{Hay09}, HE 1523-091 \citep{Fre07}
and SDSS J2357-0052 \citep{Aok10}, meet the requirement of the
extreme main r-process stars. For exploring the robustness of the
abundances in the six sample stars, we first scale the abundances of
the neutron-capture elements of CS 31082-001, CS 29497-004, HE
1219-0312, HE 1523-091 and SDSS J2357-0052  to those of CS 22892-052
and show their abundances in Figure 5 (a). The figure shows that the
abundance patterns of the five stars match the abundance pattern of
CS 22892-052  quite well within the observed uncertainties.
Furthermore, we derive the average abundance of the six stars and
scale the element abundances of the six stars to the average
abundances. In Figure 5 (b), the observed abundances of the six main
r-process stars and the average abundances, which also have been
vertically moved, are presented by filled circles and solid lines.
In Figure 5 (c), the abundance offsets of the six stars are shown.
Figure 5 (d) shows the rms abundance offsets. From the figure we
find that the abundances of the neutron-capture elements of the six
stars match the averaged abundances within 0.2 dex for the most
elements. Our finding is that not only the abundances of heavier
neutron-capture elements but also the abundances of the lighter
neutron-capture elements in the extreme main r-process stars possess
robust nature.

In Figure 6 we compare the abundances of the six weak r-process
stars with the average values of the main r-process stars. The
average values of the main r-process stars have been scaled to the
average Eu abundance of the weak r-process stars. From the figure we
can see that the abundance pattern of the neutron-capture elements
of the weak r-process stars is clearly distinct from that of the
main r-process stars, which should means that the weak r-process and
the main r-process are different astrophysical processes.

\section{ ABUNDANCE CHARACTERISTICS OF THE WEAK R-PROCESS AND MAIN R-PROCESS}

The abundance patterns of the metal-poor stars present significant
information to set important constrains on the r-process models and
Galactic chemical evolution. In the section, we will analyze
abundance characteristics of the weak r-process and main r-process.
Figure 7 (a) shows the ratios [Eu/Fe] of the extreme main r-process
stars and the weak r-process stars as a function of [Fe/H]. The main
r-process stars lie in the top of the figure, while the weak
r-process stars lie in the bottom of the figure. The two distinct
stellar categories are clear presented: the averaged ratios [Eu/Fe]
of the extreme main r-process stars is about 1.7 and the averaged
ratios [Eu/Fe] of the weak r-process stars is about -0.5. The
difference of the ratios [Eu/Fe] between the two category stars is
larger than 2.2 dex. Because of the strong enhancement of the
heavier neutron-capture elements, the extreme main r-process stars
deserve essential attention. On the other hand, the weak r-process
stars are important for exploring the weak r-process, since the main
r-process contributions to the abundances of these stars are very
small. Figure 7 (b) shows the ratios [Sr/Eu] of the extreme main
r-process stars and the weak r-process stars as a function of
[Fe/H]. The averaged ratios [Sr/Eu] of the extreme main r-process
stars and the weak r-process stars are about -1.2 and 0.4
respectively. The two distinct stellar categories are also clear
shown. The figures reconfirm the conclusion reported by \cite{Hon07}
that the abundance pattern of the main r-process stars is obviously
different from that of the weak r-process stars.

The observed flattened ratios [$\alpha$/Fe] for the low metallicity
mean that the $\alpha$ elements and Fe element in the metal-poor
stars dominantly originate from the primary-like yields (i.e. the
yields are independent of initial metallicity approximately) of the
massive stars. Figure 8 (a) shows the ratios [Sr,Y,Zr/Fe] of the
weak r-process stars as a function of [Fe/H]. The relative abundance
ratios [Sr,Y,Zr/Fe] appear to be independent of metallicity. These
correlations imply that the weak r-process elements and primary Fe
element are ejected from similar astrophysical objects, i.e.the
iron-group elements at the low metallicity are synthesized in the
massive stars (progenitors of SNe II), the weak r-process elements
should be produced in the SNe II. To investigate the abundance
characteristics of the weak r-process, the relations between the
abundances of the weak r-process elements and $\alpha$ elements are
important, since $\alpha$ elements are dominantly synthesized in the
massive stars and should be primary elements \citep{Woo95,Kob06}.
Figure 8 (b) shows the ratios [Sr,Y,Zr/Mg] of the weak r-process
stars as a function of [Fe/H]. The relative abundance ratios
[Sr,Y,Zr/Mg] also show the flattened trends: these strongly support
the viewpoint that the weak r-process elements are produced in the
SNe II and the yields of the weak r-process elements have the
primary-like nature. Based on the chemical evolution calculations,
\cite{Tra04} found that, for explaining the abundances of the
lighter neutron-capture elements, an additional nucleosynthesis
process : lighter element primary process (LEPP) is needed. The
observed trends of [Sr,Y,Zr/Fe] and [Sr,Y,Zr/Mg] for the weak
r-process stars should be the direct evidences that the weak
r-process elements are produced by the lighter element primary
process (LEPP) in the massive stars.

For further exploring the abundance characteristics of the lighter
neutron-capture elements in the two category stars, Figure 9 (a) and
(b) show the ratios [Sr/Zr] and [Y/Zr] of the extreme main r-process
stars and the weak r-process stars as a function of [Fe/H]
respectively. It is interesting to note that the averaged ratios
[Sr/Zr] and [Y/Zr] of the weak r-process stars are closed to the
corresponding ratios of the main r-process stars. The finding means
that the averaged yield fraction of Sr and Zr (or Y and Zr) of the
weak r-process events is close to that of the main r-process events.
\cite{Tra04} have found that, although the scatter of the abundance
ratios [Sr/Fe] of the metal-poor stars reaches about 2.0 dex, the
scatter of the abundance ratios [Sr/Zr] is smaller than 1.0 dex.
Note that, if the neutron-capture processes only contain the main
r-process and the weak r-process for the low metallicity, the
scatter of the observed abundance ratios [Sr/Zr] for the metal-poor
stars should be small. This prediction is qualitatively consistent
with the observed results for the metal-poor stars. Figure 10 (a)
and (b) show the ratios [Sr/Zr] and [Y/Zr] of the extreme main
r-process stars and the weak r-process stars as a function of
[Eu/Fe] respectively. From the figures we can see that, although the
two distinct stellar categories are clear shown, the averaged ratios
[Sr/Zr] and [Y/Zr] of the weak r-process stars are close to the
corresponding ratios of the main r-process stars.

In general, the abundances of the Solar system have significant
meanings, since they are seemed as the standard pattern. The
abundances of the r-process are obtained by subtracting the
abundances of the s-process from the abundances of the Solar system
\citep{Arl99}. Adopting the observed abundances of the two main
r-process stars and two weak r-process stars, the mixed abundances
have been used to compare with the abundances of the r-process in
the Solar system \citep{Zha10}. Furthermore, \cite{Li13,Han14} have
derived the pure abundances of the weak r-process and the main
r-process adopting iterative method. Note that, although a weak
r-process star is formed in a gas clouds which is polluted mainly by
the weak r-process material, the gas clouds should initially have
contained some main r-process elements, such as Eu. On the other
hand, although a main r-process star is formed in a gas clouds which
is polluted mainly by the main r-process material, the gas clouds
should initially have contained some weak r-process elements, such
as some lighter neutron-capture elements. Based on the averaged
abundances of the six weak r-process stars and the six main
r-process stars, adopting similar method presented by \cite{Li13},
we derive the abundances of the weak r-process and the main
r-process, which are called as the stellar-based weak r- and main
r-process abundances respectively. It is interesting to investigate
the relation between the stellar-based r-process abundances and the
Solar r-process abundances. In this case, we use the stellar-based
main r-process abundances to fit the Solar r-process abundances from
Eu to Pb. The Solar r-process abundances are taken from \cite{Arl99}
and \cite{Tra04} (for Sr-Nb). From the Figure 11 (a) we can see
that, for the most elements heavier than Ba, the stellar-based main
r-process abundances are good agreement with the Solar r-process
abundances. However, the stellar-based main r-process abundances are
lower than the Solar r-process abundances for the most lighter
neutron-capture elements. This fact, which have been found by many
works (e.g., \cite{Sne00,Hil02}), implies that the abundances of the
r-process in the Solar system cannot be interpreted by one main
r-process. Although the agreement does not extend to the lighter
neutron-capture elements, it is believed that the abundances have
been extended to the lighter neutron-capture elements (e.g., Sr, Y
and Zr) for the main r-process. Obviously, in order to explain the
abundances of the r-process in the Solar system, another unknown
primary process producing lighter neutron-capture elements is
needed. This process is called as lighter element primary process
(LEPP) \citep{Tra04} or weak r-process \citep{Wan06}.

\cite{Tra04} have reported that it is not clear how the LEPP
abundances for the lighter neutron-capture elements in low
metallicity evolve to corresponding abundances in the Solar system
and metal-rich stars. The missing component mentioned above should
meet two requirements at least. Firstly, it have primary nature and
should appear in the abundances of the metal-poor stars. Secondly,
it could compensate the abundance deficiency of the lighter
neutron-capture elements of the main r-process shown in Figure 11
(a). Obviously, new derived weak r-process component have meet the
first requirement. For exploring the role of the weak r-process
component, after adding the contribution of the weak r-process
component, we use the combined abundances of the main r-process and
the weak r-process to fit the abundances of the Solar r-process. In
Figure 11 (b), the fitted results are shown and compared with the
abundances of the Solar r-process. The comparison reveals a nearly
perfect agreement between the calculated abundances and the
abundances of the Solar r-process in the wide range of elements from
Sr to Pb. The derived results have significant meanings. Firstly,
the weak r-process component can compensate the abundance deficiency
of the lighter neutron-capture elements of the main r-process, which
implies the abundances of the Solar r-process are the combined
results of the weak r-process component and the main r-process
component. Secondly, although the names ``LEPP" and ``the weak
r-process" are different, they are one identical process, whose
abundances are the part of the abundances of the r-process. Thirdly,
the production of the weak r-process appear in the metal-poor stars
and have primary nature, which is consistent with the LEPP predicted
by \cite{Tra04}. Fourthly, although the weak r-process component and
the main r-process component are derived from the abundances of the
metal-poor stars, the abundances of the Solar r-process can be
explained by the mixing of the two components, which should be
another evidence that the abundance pattern of the weak r-process is
uniform and independent of metallicity. Fifthly, because $\alpha$
elements, iron-group elements and the weak r-process elements in the
metal-poor stars originate from the similar astrophysical objects,
while the main r-process elements do not couple with the light
elements, the two r-processes should not occurred in the similar
astrophysical objects. In this case, we suggested that the weak
r-process and the main r-process are two distinct astrophysical
processes. Finally, because the weak r-process elements are coupled
with the primary Fe and $\alpha$ elements, the weak r-process is a
general phenomenon in our Galaxy, which means that almost all the
metal-poor stars and population I stars have been polluted by the
weak r-process.

\section{CONCLUSIONS}

The elemental abundances of the metal-poor stars contain wealth
nucleosynthetic information. In this aspect, the abundances of
the weak r-process stars and extreme main r-process stars are
very important for investigating the sites and physical conditions
of the r-process. In this work, started from the abundances of the
metal-poor star HD 140283, we investigate the abundance characteristics
of the weak r-process and main r-process. Our results and concluding
remarks can be listed as follows:

1. Based on the obtained component coefficients, we find that HD
140283 is a weak r-process star. The astrophysical reason that the
observed ratio [Ba/Eu]$=-0.58\pm 0.15$ is close to the ratio of the main
r-process is that Ba and Eu in this weak r-process star mainly come
from the main r-process.

2. Through comparing of the abundances in the six weak r-process
stars, we find that the abundances of the weak r-process stars
possess uniform nature. Because the uniform nature of the weak
r-process stars has extended to the abundances of the light and
iron-group elements, the abundance pattern contains more information
and could be used to constraint astrophysical sites and conditions
of the weak r-process nucleosynthesis.

3. For the extreme main r-process stars, we find that not only the
abundances of the heavier neutron-capture elements but also the
abundances of the lighter neutron-capture elements possess robust
nature, which should provide more astrophysical information to
constraint the main r-process nucleosynthesis.

4. For the weak r-process stars, The variations of the ratios
[Sr,Y,Zr/Fe] and [Sr,Y,Zr/Mg] with metallicity [Fe/H] show the
flattened trends. Because the $\alpha$ elements and Fe element in
the metal-poor stars dominantly originate from the primary-like
yields of the massive stars, the correlations imply that the weak
r-process elements and primary Fe element (and primary $\alpha$
elements) are ejected from similar astrophysical objects. The
observed trends of [Sr,Y,Zr/Fe] and [Sr,Y,Zr/Mg] for the weak
r-process stars should be the direct evidences that the weak
r-process elements are produced by the lighter element primary
process (LEPP) in the massive stars.

5. The averaged ratio [Sr/Zr] (or [Y/Zr]) of the weak r-process
stars are closed to the corresponding ratio of the main r-process
stars, which means that the averaged yield fraction of Sr and Zr (or
Y and Zr) of the weak r-process events is close to that of the main
r-process events.

6. Based on the averaged abundances of the weak r-process stars and
the main r-process stars, we derive the abundance patterns of the
weak r-process and the main r-process. Because the abundances of the
main r-process have been extended to the lighter neutron-capture
elements (e.g., Sr, Y and Zr) the abundances of the weak r-process
have been extended to the light elements and iron-group elements,
the new derived abundances of the main r-process and the weak
r-process should be used to investigate the astrophysical sites and
physical conditions of the corresponding r-processes.

7. The abundances of the main r-process are good agreement with the
abundances of the r-process of the Solar system for the elements
heavier than Ba, while the abundances of the main r-process are
lower than corresponding abundances of the r-process of the Solar
system for the most lighter neutron-capture elements.

8. For exploring the role of the weak r-process component, we use
the combined abundances of the main r-process and the weak r-process
to fit the abundances of the Solar r-process. The results show a
nearly perfect agreement between the calculated abundances and the
Solar r-process abundances in the wide range of elements from Sr to
Pb, which implies that the weak r-process component can compensate
the abundance deficiency of the lighter neutron-capture elements of
the main r-process. The metallicity of the solar system is much
higher than that of the weak r-process stars. The results mean that
the pattern of the weak r-process, which contributes the weak
r-process elements to the Galaxy from early time to the epoch of
solar system formation, is ``robust" within the observed
uncertainties, i.e. the weak r-process invariably produce the
uniform abundance pattern.

In the past few decades, a large number of works have been done to
investigate the astrophysical origin of the neutron-capture
elements. In this field, the observations of the neutron-capture
elements for metal-poor stars present a important chance to
determine the abundance patterns synthesize by individual
neutron-capture processes in the early Galaxy. Up to now, the
astrophysical sites of the r-process have not been fully confirmed.
The results derived in this work could provide the more constrains
on the study of the weak r-process and main r-process. We wish the
derived results can be used to investigate the astrophysical sites
and physical conditions of the weak r-process and main r-process.
Clearly, more observational abundances, particularly for the weak
r-process stars and main r-process stars, are significant to
identify the real sites of the r-process.

\acknowledgments

This work has been supported by the National Natural Science
Foundation of China under 11273011, U1231119, 10973006, XJPT002 of
Shijiazhuang University, the Natural Science Foundation of Hebei
Province under Grant A2011205102 and the Program for Excellent
Innovative Talents in University of Hebei Province under Grant
CPRC034.

\clearpage

\begin{figure}[t]
 \centering
 \includegraphics[width=0.88\textwidth,height=0.5\textheight]{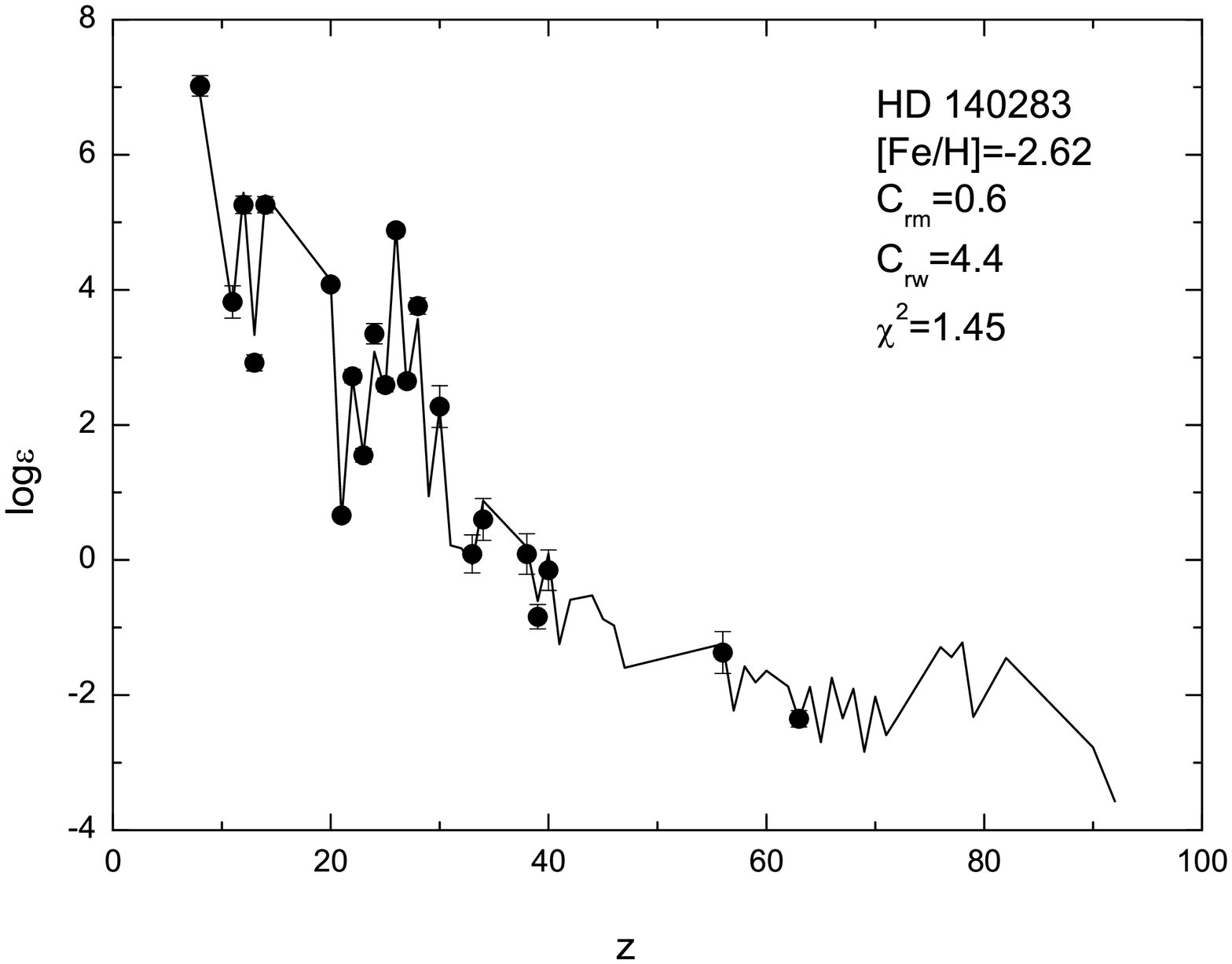}
\caption{Best fitted results of HD 140283. The filled circles
with error bars are the observed element abundances, the solid line
represents the calculated results.}
\end{figure}

\begin{figure}[t]
 \centering
 \includegraphics[width=0.88\textwidth,height=0.5\textheight]{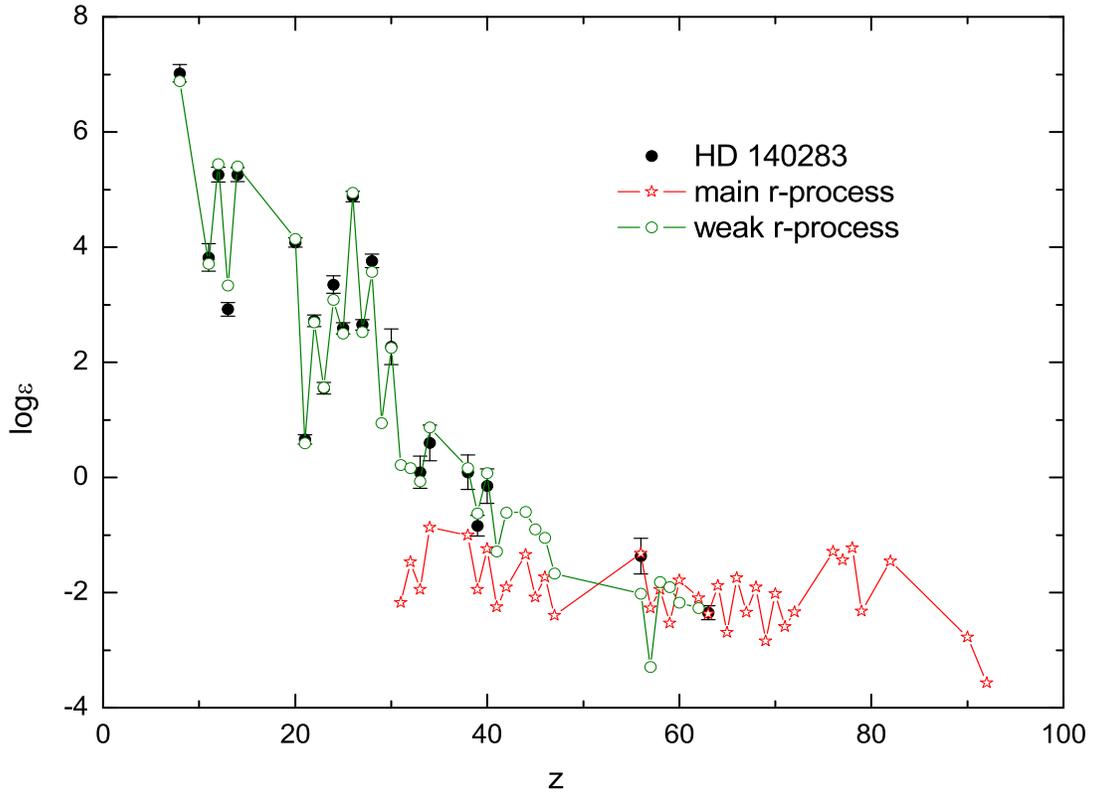}
\caption{ Comparisons of the calculated abundances of the main
r-component and weak r-component with the observed abundances. The
filled circles with error bars are the observed element abundances,
the open stars and circles represent the abundances of the main
r-component and the weak r-component, respectively. }
\end{figure}

\begin{figure}[t]
 \centering
 \includegraphics[width=0.88\textwidth,height=0.5\textheight]{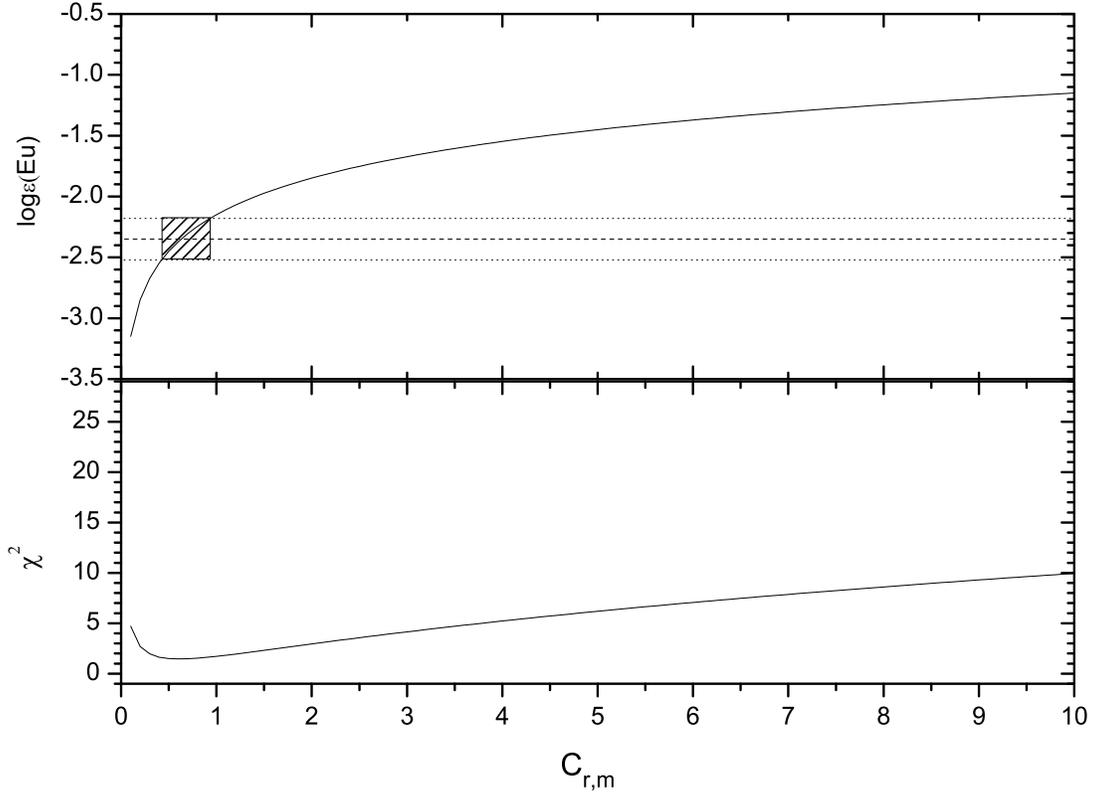}
\begin{center}
\caption{Calculated abundances $\log\varepsilon(Eu)$ (top panel) and
the reduced $\chi^{2}$ (bottom panel) as a function of the component
coefficient $C_{r,m}$, in the calculation with $C_{r,w}=0.6$. Solid
curves represent the calculated results. Dashed line represents the
observed value and dotted lines refer to the observed error. The
shaded region shows the allowed range of the component coefficient.}
\end{center}
\end{figure}

\begin{figure}[t]
 \centering
 \includegraphics[width=0.8\textwidth,height=0.88\textheight]{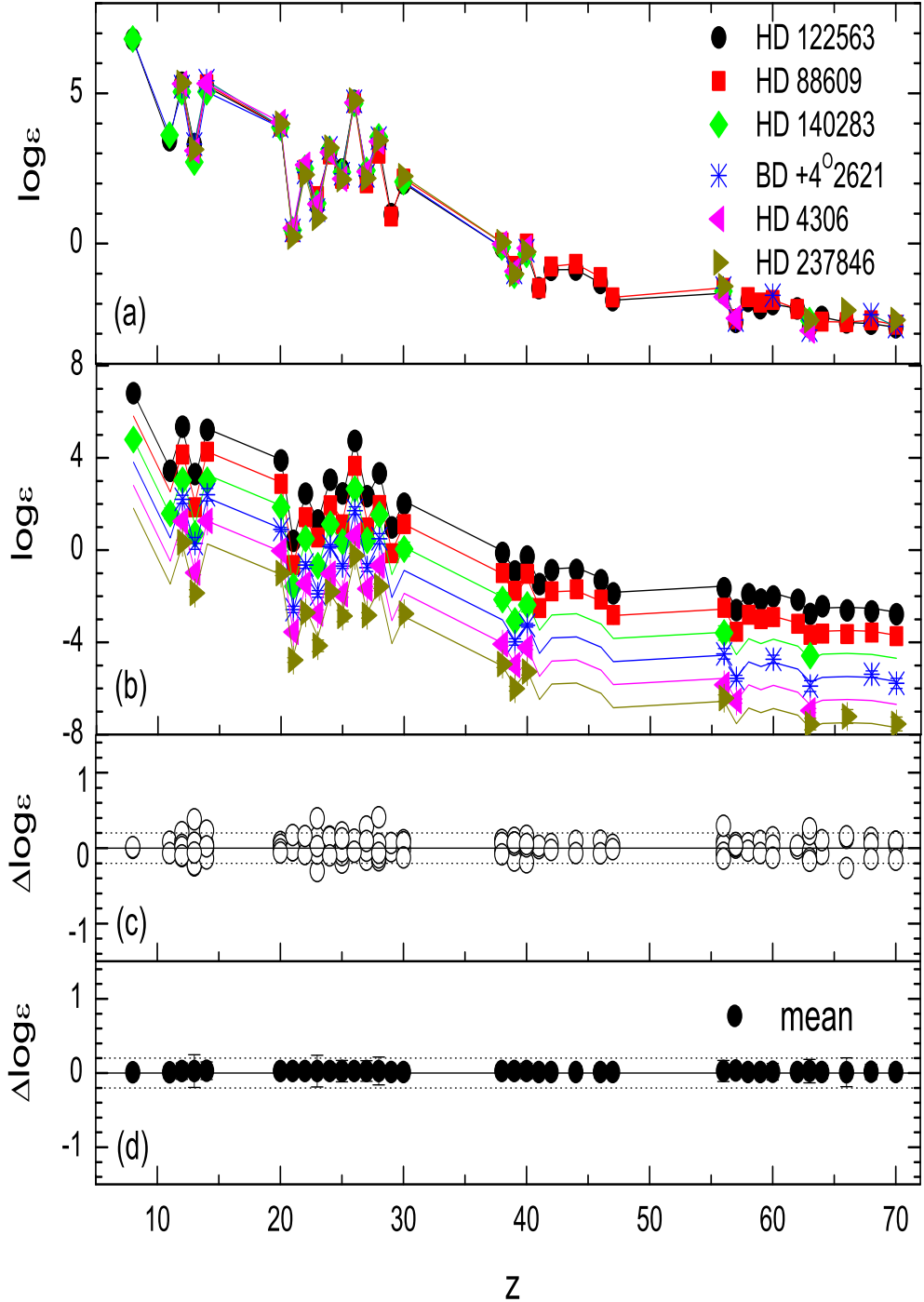}
\begin{center}
\caption{ (a): Comparisons of abundances in the five weak r-process
stars with the abundances of HD 122563. (b): Comparisons of
abundances in six weak r-process stars with the average abundance.
The observed abundances of six weak r-process stars and the average
abundances, which have been vertically moved for the sake of
display, are presented by filled circles and solid lines. (c): The
abundance offsets ($\bigtriangleup\log\varepsilon$) of the six
sample stars are shown. (d): The rms abundance offsets are shown. }
\end{center}
\end{figure}

\begin{figure}[t]
 \centering
 \includegraphics[width=0.8\textwidth,height=0.88\textheight]{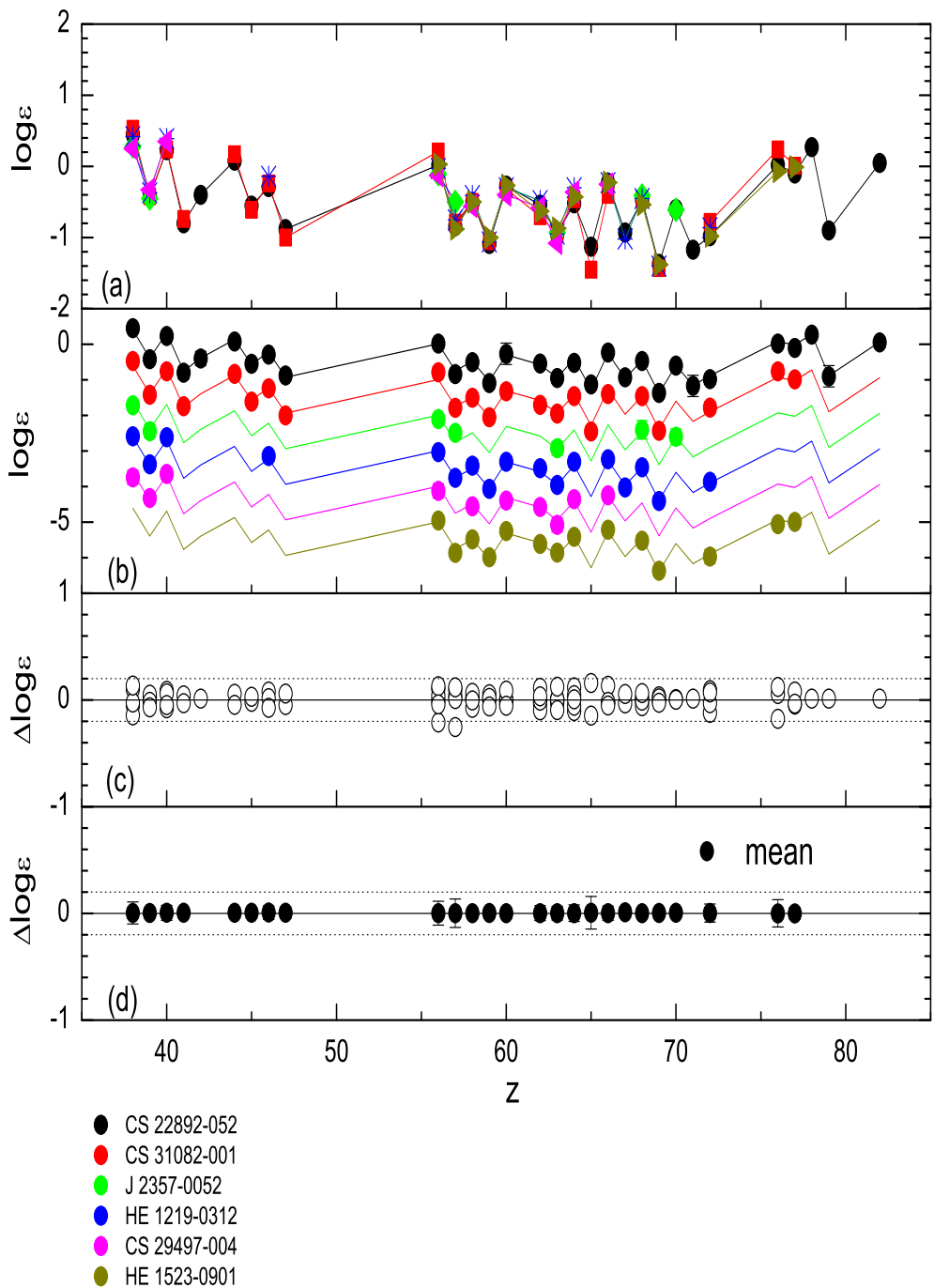}
\begin{center}
\caption{(a): Comparisons of abundances in the five main r-process stars with
the abundance of CS 22892-052. (b): Comparisons of abundances in six main r-process stars with
the average abundance. (c): The abundance offsets of the six sample stars are shown.
(d): The rms abundance offsets are shown.}
\end{center}
\end{figure}

\begin{figure}[t]
 \centering
 \includegraphics[width=0.88\textwidth,height=0.5\textheight]{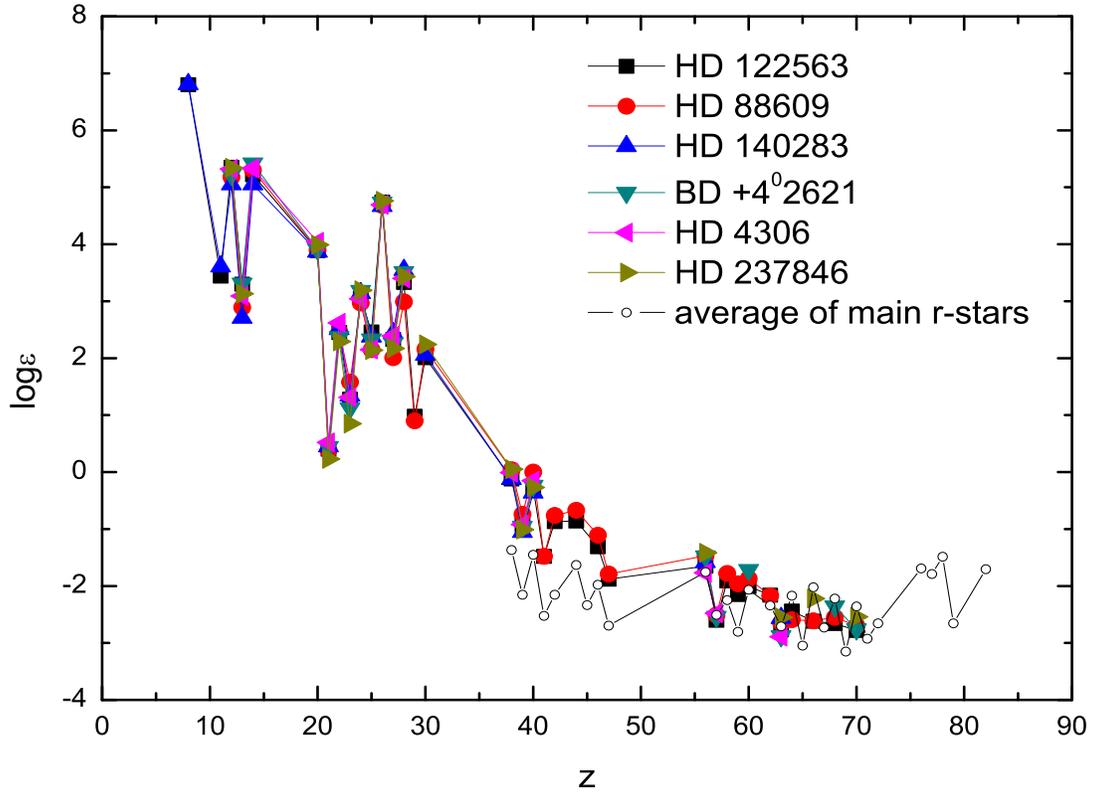}
\begin{center}
\caption{The comparison between the abundances of the six weak
r-process stars and the average values of the main r-process stars.
The average values of the main r-process stars have been scaled to
the average Eu abundance of the weak r-process stars. }
\end{center}
\end{figure}

\begin{figure}[t]
 \centering
 \includegraphics[width=1\textwidth,height=0.4\textheight]{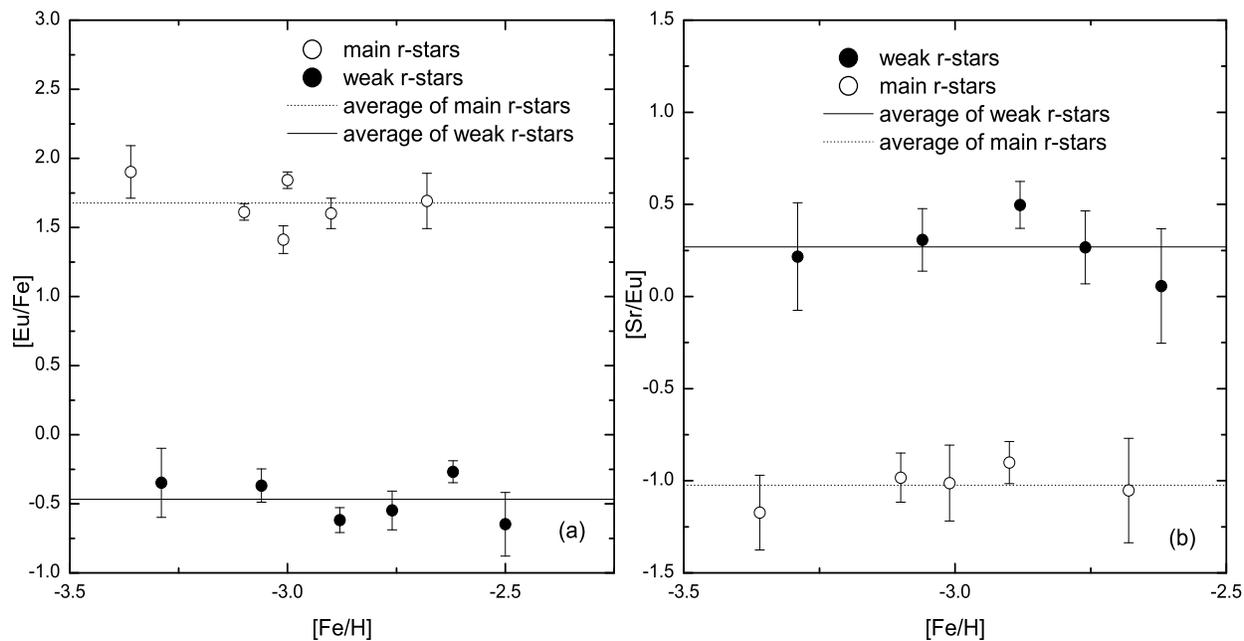}
\begin{center}
\caption{(a): The ratios [Eu/Fe] of the extreme main r-process
stars and the weak r-process stars as a function of [Fe/H]. (b): The ratios
[Sr/Eu] of the extreme main r-process stars and the weak r-process
stars as a function of [Fe/H]. The filled circles and the open circles
represent the weak r-stars and the main r-stars, respectively.
The dashed lines and the solid lines represent the averages of
the main r-stars and the weak r-stars, respectively.}
\end{center}
\end{figure}

\begin{figure}[t]
 \centering
 \includegraphics[width=1\textwidth,height=0.4\textheight]{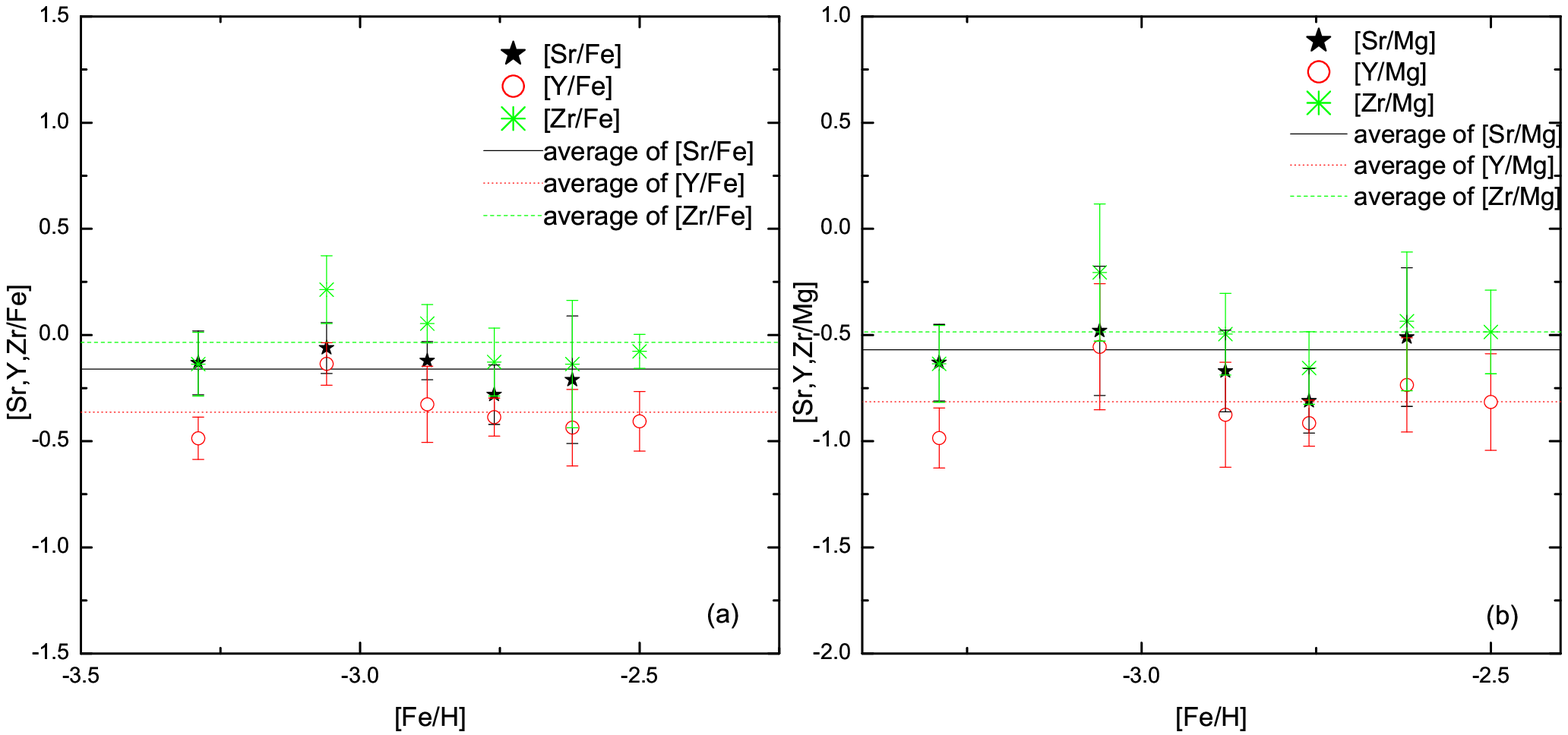}
\begin{center}
\caption{ (a): The ratios [Sr,Y,Zr/Fe] of the weak r-process
stars as a function of [Fe/H]. The stars, open circles and asterisks
represent the ratios [Sr/Fe], [Y/Fe] and [Zr/Fe]. The solid lines,
dotted lines and dashed lines represent the averages of the ratios
[Sr/Fe], [Y/Fe] and [Zr/Fe]. (b): The ratios [Sr,Y,Zr/Mg] of the weak
r-process stars as a function of [Fe/H]. The stars, open circles and
asterisks represent the ratios [Sr/Mg], [Y/Mg] and [Zr/Mg].
The solid lines, dotted lines and dashed lines represent
the averages of the ratios [Sr/Mg], [Y/Mg] and [Zr/Mg].}
\end{center}
\end{figure}

\begin{figure}[t]
 \centering
 \includegraphics[width=1\textwidth,height=0.4\textheight]{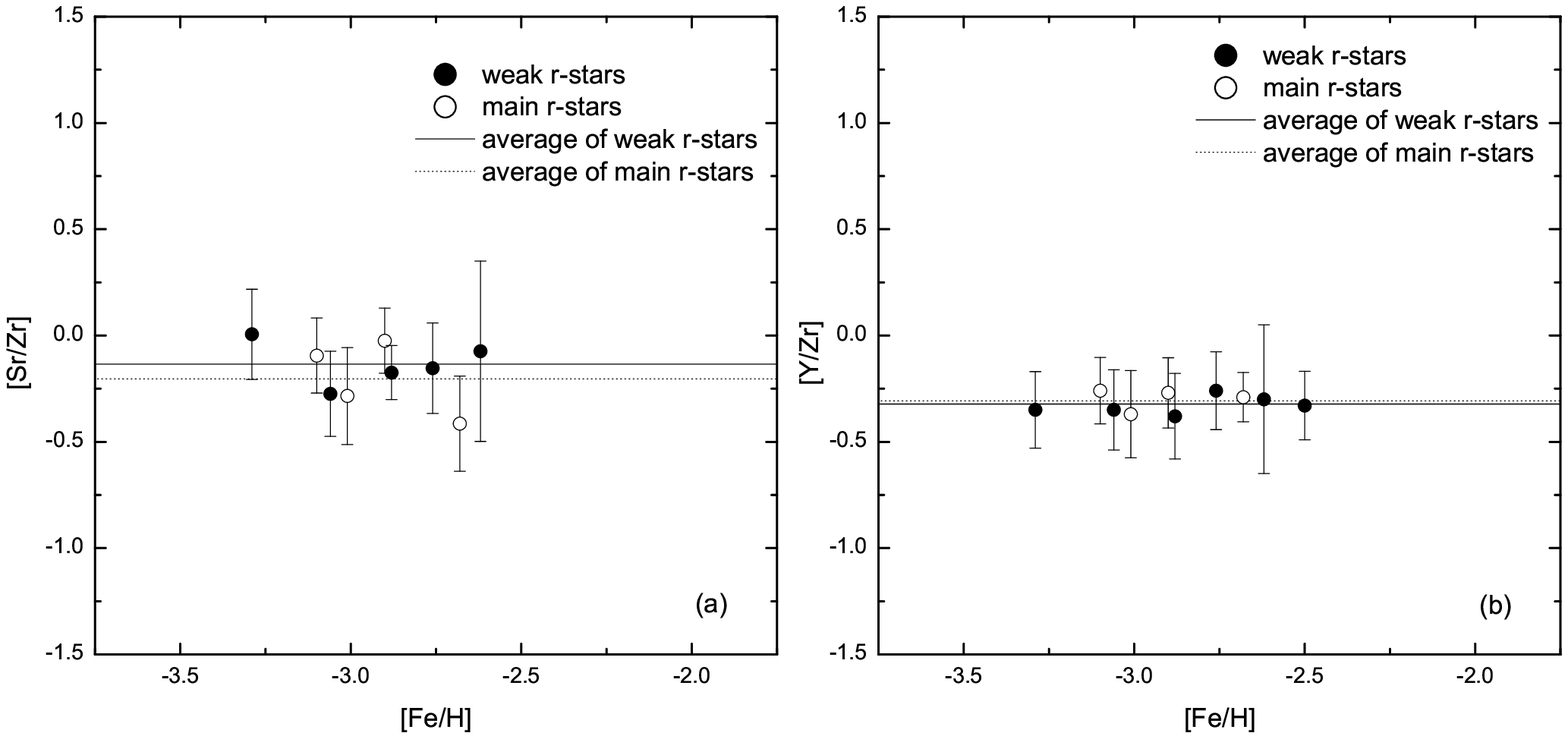}
\begin{center}
\caption{ The ratios [Sr/Zr] (a) and [Y/Zr] (b) of the
extreme main r-process stars and the weak r-process stars as a
function of [Fe/H] respectively. The filled circles and the open circles
represent the weak r-stars and the main r-stars, respectively.
The solid lines and the dashed lines represent the averages of
the weak r-stars and the main r-stars, respectively. }
\end{center}
\end{figure}

\begin{figure}[t]
 \centering
 \includegraphics[width=1\textwidth,height=0.4\textheight]{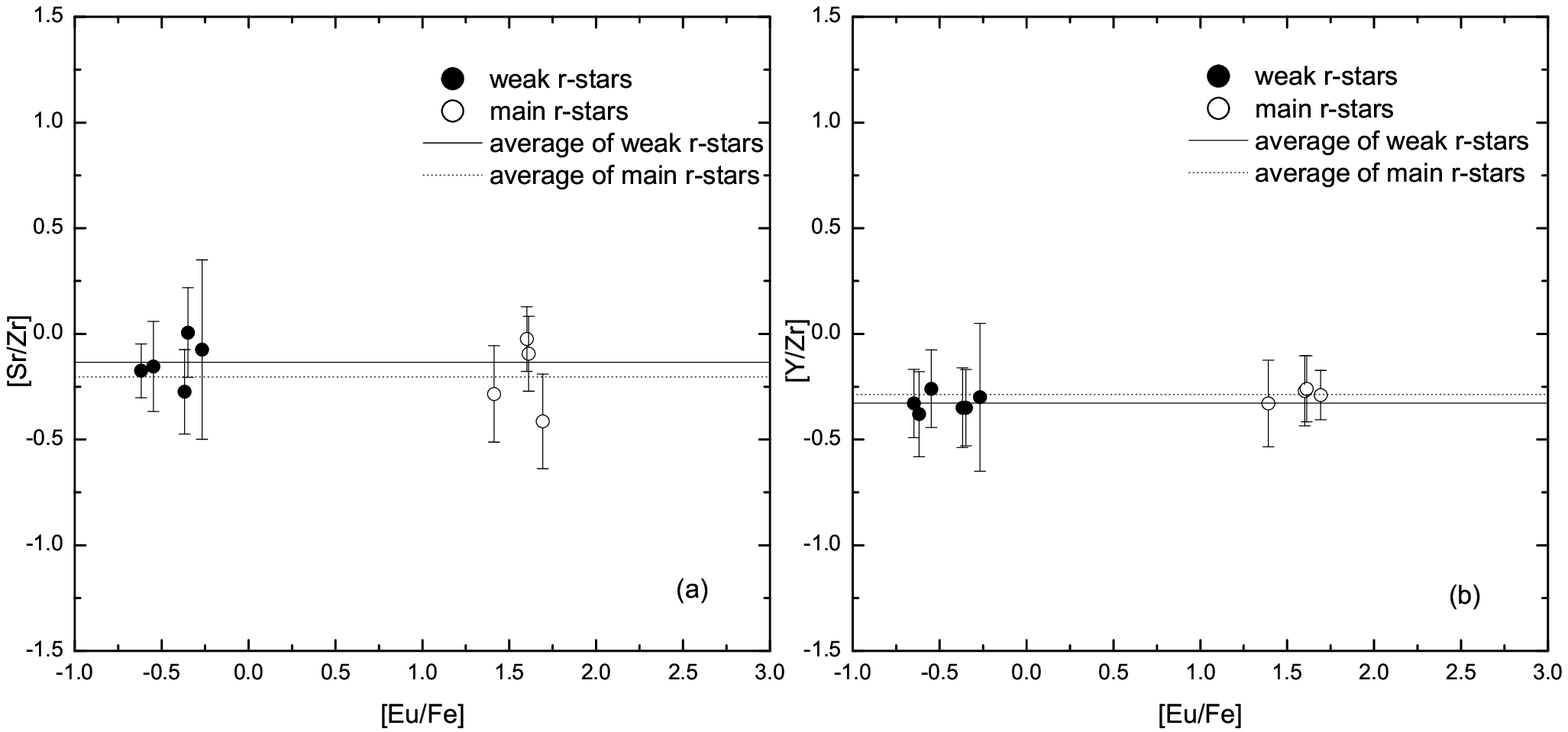}
\begin{center}
\caption{The ratios [Sr/Zr] (a) and [Y/Zr] (b) of the extreme main r-process
stars and the weak r-process stars as a function of [Eu/Fe] respectively.
The filled circles and the open circles
represent the weak r-stars and the main r-stars, respectively.
The solid lines and the dashed lines represent the averages of the
weak r-stars and the main r-stars, respectively.}
\end{center}
\end{figure}

\begin{figure}[t]
 \centering
 \includegraphics[width=1\textwidth,height=0.4\textheight]{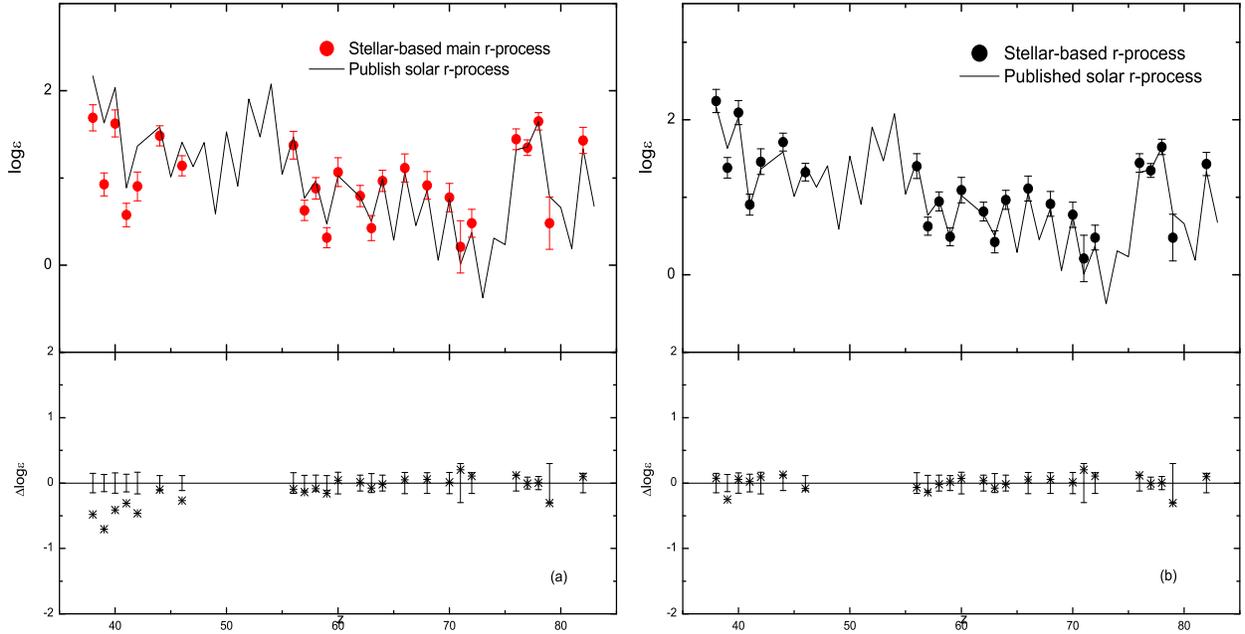}
\begin{center}
\caption{(a) Top panel: The comparison of the stellar-based main
r-process abundances (filled circles with error bars) and the Solar
r-process abundances (solid line). Bottom panel: The individual
relative offsets $[\vartriangle\log \varepsilon(X) =
\log\varepsilon(X)_{cal,mr}-\log\varepsilon(X)_{Solar,r}]$. (b) Top
panel: The comparison of combined abundances of the main r-process
and the weak r-process (filled circles with error bars) and the
abundances of the Solar r-process (solid line). Bottom panel: The
individual relative offsets $[\vartriangle\log \varepsilon(X) =
\log\varepsilon(X)_{cal,r}-\log\varepsilon(X)_{Solar,r}]$.}
\end{center}
\end{figure}



\begin{thebibliography}{}

\bibitem[Aoki et al.(2010)]{Aok10} Aoki, W., Beers, T. C., Honda, S., Carollo, D. 2010, \apj 723, L201
\bibitem[Arlandini et al.(1999)]{Arl99} Arlandini, C., K\"{a}ppeler, F., Wisshak, K., Gallino, R., Lugaro, M., Busso, M., \& Straniero, O. 1999, \apj, 525, 886
\bibitem[Barklem et al.(2005)]{Bar05} Barklem, P. S., et al. 2005, \aap, 439, 129
\bibitem[Burbidge et al.(1957)]{Bur57} Burbidge, E. M., Burbidge, G. R., Fowler, W. A., \& Hoyle, F. 1957, Rev. Mod. Phys, 29, 547
\bibitem[Bond et al.(2013)]{Bon13} Bond, H. E., Nelan, E. P., VandenBerg, D. A., Schaefer, G. H., Harmer, D. 2013, \apj, 765, L12
\bibitem[Chamberlain \& Aller(1951)]{Cha51} Chamberlain, J. W., \& Aller, L. H. 1951, ApJ, 114, 52
\bibitem[Cowan et al.(1999)]{Cow99} Cowan, J. J., Pfeiffer, B., Kratz, K.-L., Thielemann, F.-K., Sneden, C., Burles, S.,Tytler, D., \& Beers, T. C. 1999, \apj, 521, 194
\bibitem[Frebel et al.(2007)]{Fre07} Frebel, A., Christlieb, N., Norris, J. E., Thom, C., Beers, T. C., Rhee, J. 2007, \apj, 660, L117
\bibitem[Gallagher et al.(2010)]{Gal10} Gallagher, A. J., Ryan, S. G., Garc\'{i}a P\'{e}rez, A. E. et al. 2010, \aap, 523, A24
\bibitem[Hansen et al.(2014)]{Han14} Hansen, C. J., Montes, F., Arcones, A. 2014, arxiv:1408.4135v1
\bibitem[Hayek et al.(2009)]{Hay09} Hayek, W., et al. 2009, \aap, 504, 511
\bibitem[Hill et al.(2002)]{Hil02} Hill, V., et al. 2002, \aap, 387, 560
\bibitem[Honda et al.(2004)]{Hon04} Honda, S., Aoki, W., Kajino, T., Ando, H., Beers, T. C., Izumiura, H., Sadakane, K., Takada-Hidai, M. 2004, \apj, 607, 474
\bibitem[Honda et al.(2007)]{Hon07} Honda, S., Aoki, W., Ishimaru, Y., Wanajo, S. 2007, \apj, 666, 1189
\bibitem[Johnson(2002)]{Joh02} Johnson, J. 2002, \apjs, 139, 219
\bibitem[Kobayashi et al.(2006)]{Kob06} Kobayashi, C., Umeda, H., Nomoto, K., Tominaga, N., Ohkubo, T. 2006, \apj, 653, 1145
\bibitem[Lamb et al.(2002)]{Lam02} Lambert, D. L., \& Allende Prieto, C. 2002, MNRAS, 335, 325
\bibitem[Li et al.(2013)]{Li13} Li, H. J., Shen X.J., Liang, S., Cui, W. Y., Zhang, B. 2013, PASP, 125, 143
\bibitem[Magain(1995)]{Mag95} Magain, P. 1995, \aap, 297,686
\bibitem[Montes et al.(2007)]{Mon07} Montes, F., et al. 2007, \apj, 671, 1685
\bibitem[Roederer et al.(2010)]{Roe10} Roederer, I. U., Sneden, C., Thompson, I. B., Preston, G. W., \& Shectman, S. A. 2010, \apj, 711, 573
\bibitem[Roederer (2012)]{Roe12} Roederer, I. U. 2012, \apj, 756, 36
\bibitem[Siqueira et al.(2012)]{Siq12} Siqueira Mello, C., Barbuy, B., Spite, M., Spite, F. 2012, \aap, 548, 42
\bibitem[Sneden et al.(2000)]{Sne00} Sneden, C., Cowan, J. J., Ivans, I. I., Fuller, G. M., Burles, S.,Beers, T. C., Lawer, J. E. 2000,\apj, 533, L139
\bibitem[Sneden et al.(2003)]{Sne03} Sneden, C., Cowan, J. J., Lawler, J. E., Ivans, I. I., Burles, S., et al. 2003, \apj, 591, 936
\bibitem[Sneden et al.(2008)]{Sne08} Sneden, C., Cowan, J. J., \& Gallino, R. 2008, \araa, 46, 241
\bibitem[Travaglio et al.(2004)]{Tra04}Travaglio, C., Gallino, R., Arnone, E., Cowan, J., Jordan, F., Sneden, C. 2004, ApJ, 601, 864
\bibitem[Wanajo \& Ishimaru (2006)]{Wan06} Wanajo S., Ishimaru, Y.  2006, Nucl. Phys. A, 777, 676
\bibitem[Woosley et al.(1995)]{Woo95} Woosley, S. E., Weaver, T. A. 1995, APSS, 101, 181
\bibitem[Zhang et al.(2010)]{Zha10} Zhang, Jiang., Cui, Wenyuan., Zhang, Bo. 2010, MNRAS, 409, 1068

\end{thebibliography}
\end{document}